\documentclass[aps,prd,amsmath,amssymb,12pt]{revtex4}
\usepackage{graphicx}
\usepackage{bm}

\def\nonu{\nonumber}

\def\LK{\left(}
\def\RK{\right)}
\def\LBK{\left\lbrack}
\def\RBK{\right\rbrack}
\def\LB{\left\lbrace}

\def\RB{\right\rbrace}

\def\Eq#1{(\ref{#1})}

\def\sp#1#2{\left( #1 \vert  #2 \right)}
\def\lv#1{\left( #1 \right\vert}
\def\rv#1{\left\vert  #1 \right)}

\def\DL{{\cal L}}
\def\DQ{{\cal Q}}
\def\DP{{\cal P}}

\def\Tr{{\rm Tr\, }}

\def\be{\begin{equation}}
\def\ee{\end{equation}}
\def\bea{\begin{eqnarray}}
\def\eea{\end{eqnarray}}
\def\bef{\begin{figure}}
\def\ef{\end{figure}}

\def\emph#1{{\bf{#1}}}

\def\Tr{\mathop{\rm Tr}}

\def\journal#1#2#3#4{{#1} {\bf #2}, #3 (#4)}

\begin{document}
\title{Equation of Motion for Open Quantum Systems incorporating Memory and Initial Correlations}
\author{Martin Jan{\ss}en}
\email{mj@thp.uni-koeln.de}
\affiliation{Institut~f\"ur~Theoretische~Physik, Universit\"at~zu~K\"oln\\ Z\"ulpicher~Str.~77, D-50937~K\"oln}
\date{\today}
\begin{abstract}
An equation of motion for open quantum systems incorporating memory effects and initial correlations with the environment is presented in terms of an effective Liouville operator that solely acts on states of the system. 
The environment can induce memory effects via the frequency dependence of the effective Liouville and initial correlations can be mapped to a shifted frequency dependent initial state within the system. The equation of motion generalizes the well known semi-group dynamic equations.  In  generic systems the effective Liouville has a non-degenerate zero mode. By  probability conservation one can demonstrate that a generic open system reaches, in the long time limit, a  stationary state, which is  independent of any initial condition.
\end{abstract}
\maketitle

\section{Introduction}
We extend the approach to open quantum systems \cite{Fano,JanGen,Jan2017} by an effective Liouville to systems with initial correlations with the environment. 
A system coupled to an environment  forming a total closed quantum system with a von Neumann dynamics is the common framework to study  open quantum systems (see ~\cite{BreuPet,BreuerEtal} for an introduction).
  If one  neglects the entanglement of the density operator between system and environment at some single moment, one can use this as a separated initial condition to gain a closed dynamic description for the density operator reduced to the system. 
This dynamic is known (see e.g.~Sec.~II A.1 in \cite{BreuerEtal}) to be represented as  a completely positive trace preserving map, denoted as  quantum dynamical map. When memory effects can be neglected the quantum dynamical map can be
 generated by a Liouville operator $\DL$  as a semi-group dynamics  (see~\cite{AlickiQDS} for an introduction),
\be
\rho(t)=e^{-i \DL t} \rho_0\, , \label{1.3sg}
\ee
 where $\rho_0$ is the system's initial density operator which is assumed to be uncorrelated with the environment.
In \cite{JanGen} the present author has picked up the projector technique by Zwanzig \cite{Zwan} and used the Laplace transformed density operator,
\be
\rho(z):=\int_0^\infty dt \, e^{izt} \rho (t)\, , \label{1.3d}
\ee
 to get an equation of motion in frequency space, $z=\omega+i0+$, for the case of uncorrelated initial states. 
Such an approach was undertaken already by Fano in \cite{Fano} shortly after the pioneering work on the projector technique by Zwanzig \cite{Zwan}.  
With the resolvent $\LBK z-\DL(z) \RBK^{-1}$ of a frequency dependent effective Liouville  operator $\DL(z)$ the equation of motion reads \cite{JanGen} .
\be
	\rho(z) = i \LBK z-\DL(z) \RBK^{-1} \rho_0 \, , \label{1.3}
\ee
where $\rho_0$ is the system's initial density operator which is still assumed to be uncorrelated with the environment. The effective Liouville $\DL(z)$ is a frequency dependent super operator acting solely on the system. Equation~(\ref{1.3}) is a powerful   
substitute when   the semi-group structure of \Eq{1.3sg}  is missing due to memory effects. Equation~(\ref{1.3}) generates a dynamical quantum map, $\rho_0 \to \rho(t)$, by inverse Laplace transformation. 
In \cite{Jan2017} a spectral analysis of the effective Liouville operator was used to reach conclusions about general aspects of open systems. 

As far as the author knows, the incorporation of initial correlations has not yet reached a conclusive state (see \cite{BreuerEtal}).
In this note we show in Sec.~2 how to incorporate the initial correlations by the  projector and  Laplace techniques used in \cite{Fano,JanGen,Jan2017}.
Equation~(\ref{1.3}) generalizes in a simple manner to
\be
	\rho(z) = i \LBK z-\DL(z) \RBK^{-1} \rho_0(z)  \, . \label{1.3n}
\ee
 Only the initial state $\rho_0$ becomes shifted to a frequency dependent initial state $\rho(z)=\rho_0+{\Delta\rho}^{\rm corr}_0(z)$ which includes the initial correlations with the environment in an explicit manner within the formalism. The time dependent density operator is given by inverse Laplace transformation, 
\be
\rho(t) = \frac{i}{2\pi} \int_{-\infty + i0+}^{\infty+i0+}  dz \, e^{-izt} \LBK z-\DL(z) \RBK^{-1} \rho_0(z)  \, . \label{1.3t}
\ee
Although there is no  quantum dynamical map from the initial $\rho_0$ to $\rho(t)$ within the open system in a strict sense (the part $\Delta\rho^{\rm corr}_0(z)$   depends on initial properties of the environment),  the  generated dynamics \Eq{1.3t} is as operative as a semi-group dynamics $\rho(t)=e^{-i\DL t}\rho_0$ to which it reduces when the initial correlations and memory effects can be neglected.

The frequency dependent part of the initial state in \Eq{1.3n} is  generated in a similar manner as the frequency dependent part of the effective Liouville. The explicit expressions are given in Sec.~2.
In Sec.~3 we will show that  \Eq{1.3n}  allows to extend  the conclusion about the long time limit in generic open systems of \cite{Jan2017}  to systems with arbitrary initial correlations: In  generic open systems  a  stationary state is reached in the long time limit being  independent of initial conditions.
 On the other hand it opens a way to design non-generic open systems such that initial conditions can influence the final stationary states (see \cite{Albert2018} for systems with semi-group dynamics).

\section{The equation of motion}
We consider a system coupled to an environment 
and follow the quite general projector formalism initiated by \cite{Naka} and \cite{Zwan}. 
The  reduced density operator \index{reduced density operator} is defined by a projector $\DP$ on the Hilbert-Schmidt space of linear operators over the Hilbert space of the total system (simply denoted as $\DP$-space),
\be
\rho:= \DP \rho_{\rm tot} \, .\label{4.50}
\ee
Here $ \rho_{\rm tot}$ is the total density operator of the total closed system. 
As an example  consider  the  standard projector $\DP$  defined via  a partial trace over the Hilbert space of  environmental  variables (denoted by subscript$E$),
\be
\DP \cdot = \Tr_{E} (\cdot) \otimes \rho_E \, , 
\ee
where $\rho_E$ is the would-be stationary density operator of the environment -  if the environment was not coupled to the system.
The reduced density operator $\rho$ is an element of the  $\DP$-space and can  be represented by a $d\times d$ dimensional density matrix when the closed system's Hilbert space is $d$-dimensional.
The projector on the complement to the $\DP$-space is denoted as $\DQ:=1-\DP$. Both projectors fulfill the projector property ($\DP^2=\DP$, $\DQ^2=\DQ$) and the complement property ($\DP\DQ=\DQ\DP=0$.) The complement space is simply denoted as $\DQ$-space.

The phrase  environment is not restricted to spatially external variables interacting by  scattering processes with  the system but may also mean variables spatially  within the system which  will (or cannot) be treated as relevant. 
The reduced density operator for the open system will  simply be called density operator when a misunderstanding is not to be expected.

The dynamics of the total density operator is described by  a von Neumann equation  with total Hamiltonian $H_{\rm tot}$,
\be
\partial_t \rho_{\rm tot} = -i\DL_{\rm tot} \rho_{\rm tot} \label{4.52}\, ,
\ee
with the  total system's {Liouville operator} $\DL_{\rm tot}\cdot = \LBK H_{\rm tot}, \cdot\RBK$ as the generator of the closed quantum dynamics. As a super operator it is hermitian and has a  real valued spectrum $\omega_{\alpha \beta}=\epsilon_\alpha-\epsilon_\beta$, where $\epsilon_\alpha$ are the energy eigenvalues  of the total Hamiltonian.

The  solution of the closed quantum dynamics is the time evolution of the total density operator,
\be
	\rho_{\rm tot} (t) = e^{-i\DL_{\rm tot} t} {\rho_{\rm tot}}_0 \, , \label{4.53}
\ee
Equation~(\ref{4.53}) has the group-property as the unitary time evolution forms a group with inverse elements being within the group. This formalizes the reversible character of closed quantum systems.
We like to construct the dynamic equation for the reduced density operator $\rho(t)$ by using the decomposition,
\be
 \rho_{\rm tot} =\rho+ {\Delta\rho}^{\rm corr} \, , \; {\Delta\rho}^{\rm corr}:=\DQ\rho_{\rm tot} \, ,\label{4.51}
\ee
where ${\Delta\rho}^{\rm corr}$ captures correlations between system and environment. To capture memory effects we consider the spectral content of $\rho_{\rm tot} (t)$ as with \Eq{1.3d}. The group-property \Eq{4.53} of the total system transforms to the resolvent equation of the total system
\be
	 \rho_{\rm tot}(z) = i \LBK z-\DL_{\rm tot} \RBK^{-1}  {\rho_{\rm tot}}_0 \, .\label{44.55}
\ee
The resolvent $\LBK z-\DL_{\rm tot} \RBK^{-1}$  of the total system's Liouville  is  analytic in the upper frequency plane $z=\omega + i\epsilon$ and singular at the real valued spectrum of eigenvalues,  $z=\omega_{\alpha \beta}$. 

To derive an equation of motion for the reduced density operator we use the projection operators and algebraic identities for resolvent operators (pioneered by \cite{WeisWig})
 \bea
\rho(z) &=& i\DP  \LBK z-\DL_{\rm tot} \RBK^{-1} (\DP+\DQ)  {\rho_{\rm tot}}_0  \nonu \\ 
&=& i \DP \LBK z-\DL_{\rm tot} \RBK^{-1} \DP \rho_0 + i\DP \LBK z-\DL_{\rm tot} \RBK^{-1} {\DQ \Delta\rho}^{\rm corr}_0\label{444.rho} \, .
\eea
To reach an equation of motion for the  density operator $\rho$ we have to find expressions for the projected resolvent in terms of operations  living on either $\DP$-space or $\DQ$-space.
For this purpose we decompose the total system's Liouville into 
\be
	\DL_{\rm tot} = \DL_\DP + \DL_{\DP\DQ} + \DL_{\DQ\DP} +\DL_\DQ \, , \label{44.Li}
\ee
and use the algebraic identities
\bea
 [A-B]^{-1} &=& A^{-1} + A^{-1} B  [A-B]^{-1} \, , \label{444.58a}  \\ 
 {[A-B]}^{-1} &=& A^{-1} +   [A-B]^{-1} B A^{-1} \,  , \label{444.58b}
\eea
which can be verified by multiplication with $A-B$ from the right or from the left, respectively.
Therefore we can write for the projected total resolvent
\bea
\DP{\LBK z- \DL_{\rm tot} \RBK^{-1}}\DP &=&   \LBK z- \DL_\DP \RBK^{-1} + \nonu\\
&+& \LBK z- \DL_\DP \RBK^{-1} \LK \DL_\DQ +\DL_{\DP\DQ} + \DL_{\DQ\DP} \RK \LBK z- \DL_{\rm tot} \RBK^{-1}\DP \, , \label{444.61} \\ 
\DQ{\LBK z- \DL_{\rm tot} \RBK^{-1}}{\DP} &=&   \DQ\LBK z- \DL_\DQ \RBK^{-1}\DP +  \nonu \\
&+& \LBK z- \DL_\DQ \RBK^{-1} \LK \DL_\DP +\DL_{\DP\DQ} + \DL_{\DQ\DP} \RK \LBK z- \DL_{\rm tot} \RBK^{-1}\DP \, , \label{444.62} \\
\DP{\LBK z- \DL_{\rm tot} \RBK^{-1}}{\DQ} &=&   \DP \LBK z- \DL_\DQ \RBK^{-1}\DQ +  \nonu \\
&+& \DP \LBK z- \DL_{\rm tot} \RBK^{-1} \LK \DL_\DP +\DL_{\DP\DQ} + \DL_{\DQ\DP} \RK \LBK z- \DL_\DQ \RBK^{-1} \, . \label{444.63} 
\eea
Due to the complementary character of projectors we conclude
\bea
\DP{\LBK z- \DL_{\rm tot} \RBK^{-1}}\DP &=&   \LBK z- \DL_\DP \RBK^{-1} +\nonu \\
&+&  \LBK z- \DL_\DP \RBK^{-1} \LK \DL_{\DP\DQ} \LBK z- \DL_\DQ \RBK^{-1} \DL_{\DQ\DP} \RK \LBK z- \DL_{\rm tot} \RBK^{-1}\DP \, , \label{444.64} \\
\DP{\LBK z- \DL_{\rm tot} \RBK^{-1}}\DQ &=&    \DP  \LBK z- \DL_{\rm tot} \RBK^{-1} \DL_{\DP\DQ} \LBK z- \DL_\DQ \RBK^{-1} \, .\label{444.65}
\eea
Equation~(\ref{444.64}) tells that the total resolvent projected to $\DP$-space can be written as the resolvent of an effective Liouville operating solely on states of the open system (i.e.~on $\DP$-space),
\be
\DP{\LBK z- \DL_{\rm tot} \RBK^{-1}}\DP= {\LBK z- \DL(z) \RBK^{-1}} \, , \label{444.66}
\ee
where the
 effective Liouville  reads 
\be	
\DL(z) =  \DL_{\DP} + \DL_{\DP\DQ} \LBK z- \DL_\DQ \RBK^{-1} \DL_{\DQ\DP} \, .\label{4.57}
\ee
The structure is formally the same as for effective Hamiltonians (see e.g. \cite{VolZel} and references therein) and has  the  interpretation: The first term $\DL_{\DP}$ is the closed system's Liouville in the absence of an environment and the second term
describes virtual processes in the system triggered by the environment, $\DL_{\DP\DQ} \LBK z- \DL_\DQ \RBK^{-1} \DL_{\DQ\DP}$, hopping to $\DQ$-space, there taking a lift with isolated $\DQ$-propagator and finally hopping back to $\DP$-space.

The effective Liouville can be used as well in \Eq{444.65} and we find
\be
\DP{\LBK z- \DL_{\rm tot} \RBK^{-1}}\DQ =      \LBK z- \DL(z) \RBK^{-1} \DL_{\DP\DQ} \LBK z- \DL_\DQ \RBK^{-1} \, . \label{444.67}
\ee
With \Eq{444.66} and \Eq{444.67} we can rewrite \Eq{444.rho} 
\be
	\rho(z) = i \LBK z-\DL(z) \RBK^{-1} \LK \rho_0 + {\Delta\rho}^{\rm corr}_0(z) \RK   \label{444.rho2} \, ,
\ee
with 
\be
	{\Delta\rho}^{\rm corr}_0(z)=  \DL_{\DP\DQ} \LBK z- \DL_\DQ \RBK^{-1} {\Delta \rho}^{\rm corr}_0 \label{444.rho3}
\ee
as a virtual change of initial state within the system, caused by the initial correlation ($\DQ{\rho_{\rm tot}}_0$) that gets a lift by the isolated $\DQ$-propagator and hops to $\DP$-space.

Equation~(\ref{444.rho2}) is the result announced in \Eq{1.3n}.  It is an equation of motion defined  solely for states of the system. The environment enters in an operative way through the couplings, $\DL_{\DP\DQ}, \DL_{\DQ\DP}$, and the isolated $\DQ$-propagator, $\LBK z- \DL_\DQ \RBK^{-1}$.

As discussed in \cite{Jan2017} the effective Liouville is  no longer hermitian. Provided the system's energy spectrum can be considered as  relative discrete with respect to the  environment's dense continous spectrum,  the $\DQ$-propagator leads to  non-positive imaginary contributions to the eigenvalues of  the effective Liouville, indicating that the system undergoes  a dynamic phase transition to irreversibility with relaxation and decoherence when coupled to an environment.

Before we  will discuss the generic behavior of relaxation to  a single stationary state, we heuristically estimate the importance of initial correlations at intermediate times.
The couplings $\DL_{\DP\DQ}, \DL_{\DQ\DP}$ give rates for transitions between $\DP$- and $\DQ$-spaces and thus  set inverse time scales, denoted as $1/t_{PQ}$. The propagator sets (for each frequency $z$) a time scale of propagation in $\DQ$-space, denoted as $t^z_Q$.
Thus, by comparing these time scales involved in \Eq{444.rho3} we may expect that initial correlations are not important as long as $t_Q\ll t^z_{QP}$,  otherwise they will be important.
\section{The long time limit}
The Laplace transformed density of states $\rho(z)$ allows for an easy analysis of the long time limit in the usual sense as a long time average, defined by
\be
f_\infty=\lim_{\epsilon \to 0+} \epsilon \int_0^\infty f(t) e^{-\epsilon t} dt \, 
\ee for a quite arbitrary time dependent function $f(t)$.  Any oscillations that could still be present at large times in $f(t)$ will be averaged out.
In the same sense it holds true that the long time limit is  given by
\be
	f_\infty=\lim_{z\to 0} -izf(z) \, \label{3.1}
\ee
where $f(z)$ is the Laplace transformed of $f(t)$ (in analogy to \Eq{1.3d}).

We take this limit on the equation of motion \Eq{1.3n}
\be
\rho_\infty = \lim_{z\to 0} z \LBK z-\DL(z) \RBK^{-1} \LK \rho_0 + {\Delta\rho}^{\rm corr}_0(z) \RK \label{3.2}
\ee
and exclude that ${\Delta\rho}^{\rm corr}_0(z)$ may accidentally  have a singular behavior for $z\to 0+$. 
The long time limit is then determined by the zero limit of $z/(z-\DL(z))$. This limit is non-vanishing due to the existence of zero modes of $\DL(0+)$.
Let us denote the projector on the space of zero modes of $\DL(0+)$ by $\Pi^0_{\DL(0+)}$, then the long time limit of the density operator reads
\be
	\rho_\infty = \Pi^0_{\DL(0+)} \LK \rho_0 + {\Delta\rho}^{\rm corr}_0 (0+) \RK \, . \label{3.3}
\ee
That the effective Liouville must have a zero mode is a consequence of probability conservation which means $\Tr \rho(z)=i/z$ and $\Tr \DL(z)=0$ for every $z$.
Thus, the effective Liouville has the unit matrix $1$ as a  left eigenmatrix with eigenvalue $0$ and, consequently, there must exist a right eigenmatrix with eigenvalue $0$, too.
In generic systems the zero mode will be non-degenerate and by normalizing the right zero mode to unit trace we can write the projector on the zero mode as
\be
\Pi^0_{\DL(0+)}= \rv{\rho_\infty}\lv{1}\, , \label{3.4}
\ee
where the notation refers to the Hilbert-Schmidt scalar product ($\sp{A}{B}:=\Tr \LB A^\dagger B\RB$).
Since the scalar product of the left unit matrix with the initial state amounts to taking the trace, any influence of the initial state gets lost. This substantiates our statement of the introduction: 
In  generic open systems  a  stationary state is reached in the long time limit being  independent of initial conditions, even if strong initial correlations with the environment had been present.
The heuristic estimate for the relaxation time $\tau$ after which the long time limit emerges is given by the virtual processes (hopping rate $1/t_{PQ}$ and propagation time $t_Q$) present in the effective Liouville and reads $\tau=t^2_{PQ} /t^{0+}_Q$.

Equation~(\ref{3.4}) will change to a sum of projectors on different eigenstate combinations of zero modes in the case of degenerate zero modes. It is obvious that in such case the scalar product of left eigenmatrices with the initial state is not simply unity and these scalar products store information about the initial state.
This  opens a way to design non-generic open systems such that initial conditions can influence the final stationary states as discussed in \cite{Albert2018} for systems with semi-group dynamics.

{\bf Acknowledgements: }
I thank J\'anos Hajdu for many enlightening discussions and Rochus Klesse for clarifying discussions. I also thank Victor~V. Albert for sharing the idea to store information about the initial state in long lived states. 

\end{document}